\documentclass[prd,showpacs, twocolumn,amsmath,amssymb]{revtex4}

\usepackage{amssymb}
\usepackage{graphicx}
\usepackage{dcolumn}
\usepackage{bm}
\usepackage{epsfig}
\usepackage{amsfonts}
\usepackage{keyval,graphicx}
\usepackage{textcomp,wasysym}
\usepackage{subfigure}
\usepackage[none]{hyphenat}
\usepackage{xcolor}

\usepackage{array}
\newcolumntype{+}{>{\global\let\currentrowstyle\relax}}
\newcolumntype{^}{>{\currentrowstyle}}

\begin{document}



\title{\boldmath Observation of two new $N^*$ resonances in $\psi(3686) \rightarrow p\bar{p}\pi^0$}

\author{\small
M.~Ablikim$^{1}$, M.~N.~Achasov$^{5}$, D.~J.~Ambrose$^{40}$,
F.~F.~An$^{1}$, Q.~An$^{41}$, Z.~H.~An$^{1}$, J.~Z.~Bai$^{1}$,
Y.~Ban$^{27}$, J.~Becker$^{2}$, N.~Berger$^{1}$
% ...
\\
\vspace{0.2cm}
(BESIII Collaboration)\\
\vspace{0.2cm} {\it
$^{1}$ Institute of High Energy Physics, Beijing 100049, P. R. China\\
$^{2}$ Bochum Ruhr-University, 44780 Bochum, Germany\\
% ...
$^{5}$ G.I. Budker Institute of Nuclear Physics SB RAS (BINP), Novosibirsk 630090, Russia\\
% ...
$^{27}$ Peking University, Beijing 100871, P. R. China\\
% ...
$^{40}$ University of Rochester, Rochester, New York 14627, USA\\
$^{41}$ University of Science and Technology of China, Hefei 230026, P. R. China\\
% ...
}}

\vspace{0.4cm}



\collaboration{BESIII Collaboration}

\date{\today}

\begin{abstract}
Based on 106$\times10^6 \psi(3686)$ events collected with the BESIII
detector at the BEPCII facility, a partial wave analysis of
$\psi(3686)\rightarrow p\bar{p}\pi^0$ is performed. The branching
fraction of this channel has been determined to be
$B(\psi(3686)\rightarrow p\bar{p}\pi^0) = (1.65\pm0.03\pm0.15)\times
10^{-4}$. In this decay, 7 $N^*$ intermediate resonances are
observed. Among these, two new resonances, $N(2300)$ and $N(2570)$
are significant, one $1/2^+$ resonance with a mass of
$2300^{+40}_{-30}$$^{+109}_{-0}$ $\mathrm{MeV}/c^2$ and width of
$340^{+30}_{-30}$$^{+110}_{-58}$ $\mathrm{MeV}/c^2$, and one $5/2^-$
resonance with a mass of $2570^{+19}_{-10}$$^{+34}_{-10}$
$\mathrm{MeV}/c^2$ and width of $250^{+14}_{-24}$$^{+69}_{-21}$
$\mathrm{MeV}/c^2$. For the remaining 5 $N^*$ intermediate
resonances ($N(1440)$, $N(1520)$, $N(1535)$, $N(1650)$ and
$N(1720)$), the analysis yields mass and width values which are
consistent with those from established resonances.
\end{abstract}

\maketitle


Although symmetric non-relativistic three-quark models of baryons
are quite successful in interpreting low-lying excited  baryon
resonances, they tend to predict far more excited states than are
found experimentally (``missing resonance
problem")~\cite{MISSING_1,MISSING_2}. From the theoretical point of
view, this could be due to a wrong choice of the degrees of freedom,
and models considering di-quarks have been proposed~\cite{diquark}.
Experimentally, the situation is very complicated due to the large
number of broad and overlapping states that are observed. Moreover,
in traditional studies using tagged photons or pion
beams~\cite{CLAS_1,CLAS_2,CLAS_3,CLAS_4,CLAS_5, ELSA_1, ELSA_2,
MAMI}, both isospin 1/2 and isospin 3/2 resonances are excited,
further complicating the analysis.

An alternative method to investigate nucleon resonances employs
decays of charmonium states such as $J/\psi$ and $\psi(3686)$. By
selecting specific decay channels, such as $\psi(3686) \rightarrow
p\bar{p}\pi^0$, $N^*$ intermediate resonances coupling to $p\pi^0$
or $\bar{p}\pi^0$ can be studied. Here, $\Delta$ resonances are
suppressed due to isospin conservation. As a consequence, the
reduced number of states greatly facilitates the
analysis~\cite{Lihb}.

$N^*$ production in $J/\psi\rightarrow p\bar{p}\eta$ was studied
using partial wave analysis at BES~\cite{Bai:2001ua}, and two $N^*$
resonances were observed. In a recent analysis of $J/\psi
\rightarrow p\bar{n}\pi^-+c.c.$~\cite{N2065PRL}, a new $N^*$
resonance around 2000 MeV/$c^2$ named $N(2065)$ was observed. This
$N(2065)$ was also observed in the decay of $J/\psi \rightarrow
p\bar{p}\pi^0$~\cite{Ablikim:2009iw}. The production of $N(2065)$ in
$J/\psi$ decays occurs close to the edge of the phase space. Thus, a
similar search for this resonance in the $\psi(3686)$ decays should
provide further insight.

In the work of the CLEO Collaboration~\cite{CLEO},
$\psi(3686)\rightarrow p\bar{p}\pi^0$ was studied using
$24.5\times10^6 \psi(3686)$ events. With the invariant mass spectra
of $p\pi^0$ and $p\bar{p}$, two $N^*$ resonances ($N(1440)$,
$N(2300)$) and two $p\bar{p}$ resonances($R_1(2100), R_2(2900)$)
were investigated without taking into account possible interferences
between the resonances. The inclusion of R(2100) is suggested by a
threshold enhancement in the $p\bar{p}$ mass spectrum. The
concentration of events below 1800 MeV/$c^2$ in the $p\pi^0$ mass
spectrum is considered as the contribution of $N(1440)$ alone.

In this Letter, we briefly report a study of $N^*$ resonances from
$\psi(3686) \rightarrow p\bar{p}\pi^0$ based on a data sample of 160
$pb^{-1}$ corresponding to 106 million $\psi(3686)$ decays collected
with the upgraded Beijing Spectrometer (BESIII), located at the
Beijing Electron-Positron Collider (BEPCII)~\cite{BESIII_BEPCII}.
The full details will be published later.

The BESIII detector is composed of a helium-gas based drift chamber
(MDC), a time-of-flight (TOF) system, a CsI (Tl) electromagnetic
calorimeter (EMC), a super-conducting solenoid magnet and a
resistive plate chambers based muon chamber. More detailed
information about the detector can be found in
Ref.~\cite{BESIII_BEPCII}.

The final state in this decay is characterized by two charged tracks
and two photons. Two charged tracks with opposite charge are
required. Each track is required to have its point of closest
approach to the beam axis within $\pm20$ cm of the interaction point
in the beam direction and within 2 cm of the beam axis in the plane
perpendicular to the beam. The polar angle of the track is required
to be within the region of $|\cos(\theta)|<$0.8.

The TOF and the specific energy loss dE/dx of a particle measured in
the MDC are combined to calculate particle identification (PID)
probabilities for pion, kaon and proton hypotheses. For each track,
the particle type yielding the largest probability is assigned. In
this analysis, one charged track is required to be identified as a
proton and the other one as an anti-proton.

Photon candidates are selected by requiring a minimum energy
deposition of 25 MeV in the barrel EMC or 50 MeV in the endcap EMC.
To reject photons due to charged particle radiation production, the
angle between the photon candidate and the proton is required to be
greater than $10^{\circ}$. A more stringent cut of $30^{\circ}$
between the photon candidate and anti-proton is applied to exclude
the large number of photons from anti-proton annihilation.

For events with one proton, one anti-proton and at least two
photons, a kinematic fit (4C) with the sum of four-momenta of all
particles constrained to the energy and three momentum-components of
the initial $e^+e^-$ system is applied. A further kinematic fit (5C)
with one more constraint of $\pi^0$ mass for the two photons is
applied to provide more accurate momentum information of the final
states. When more than two photons are found in a candidate event,
all possible $p\bar{p}\gamma\gamma$ combinations are considered and
the one yielding the smallest $\chi^2_{5C}$ is retained for further
analysis.

\begin{figure}[htbp]
   \centerline{
   \psfig{file=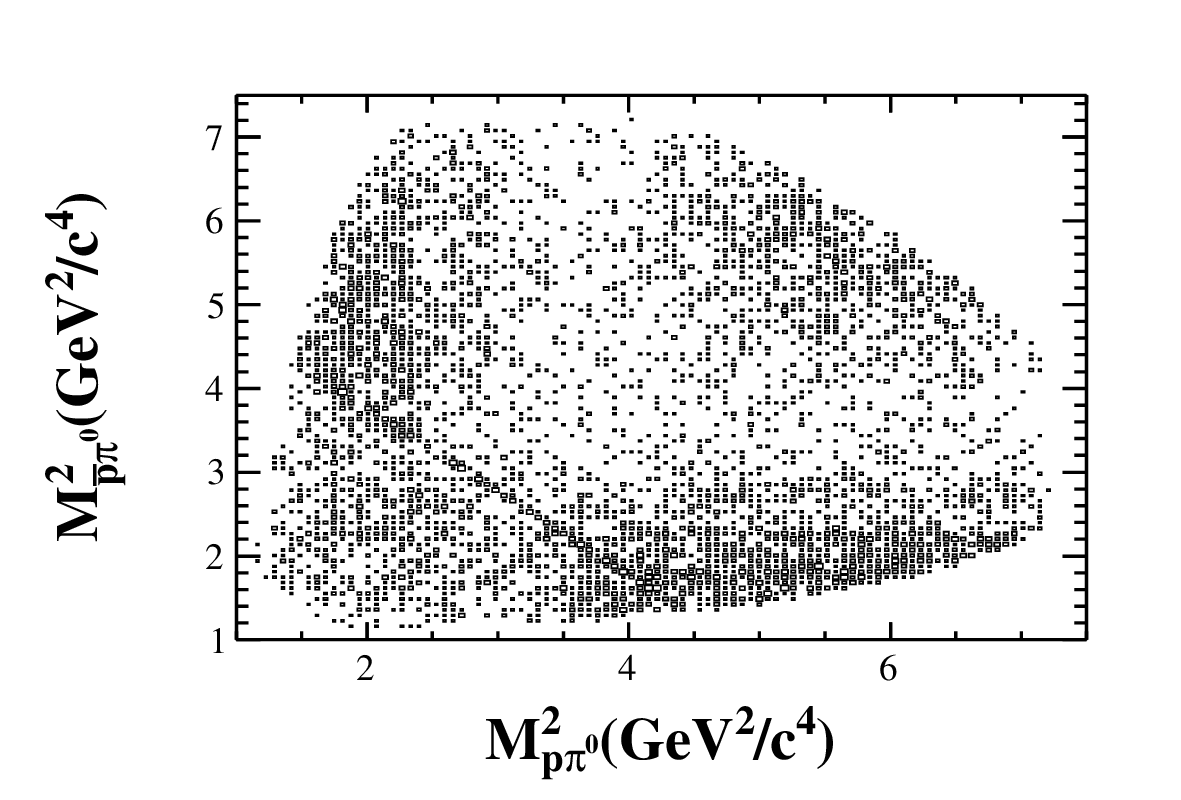,width=4.6cm, angle=0}
              \put(-27,67){(a)}
   \psfig{file=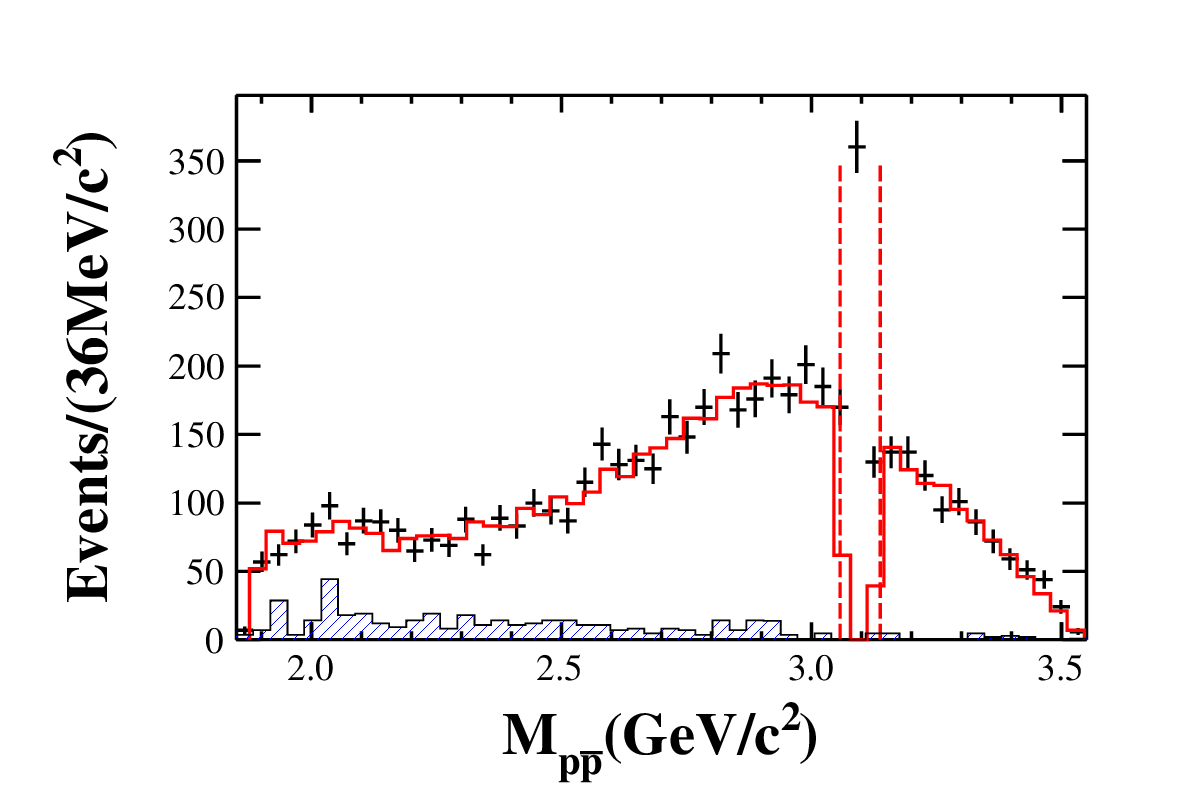,width=4.6cm, angle=0}
              \put(-27,67){(b)}}
   \centerline{
   \psfig{file=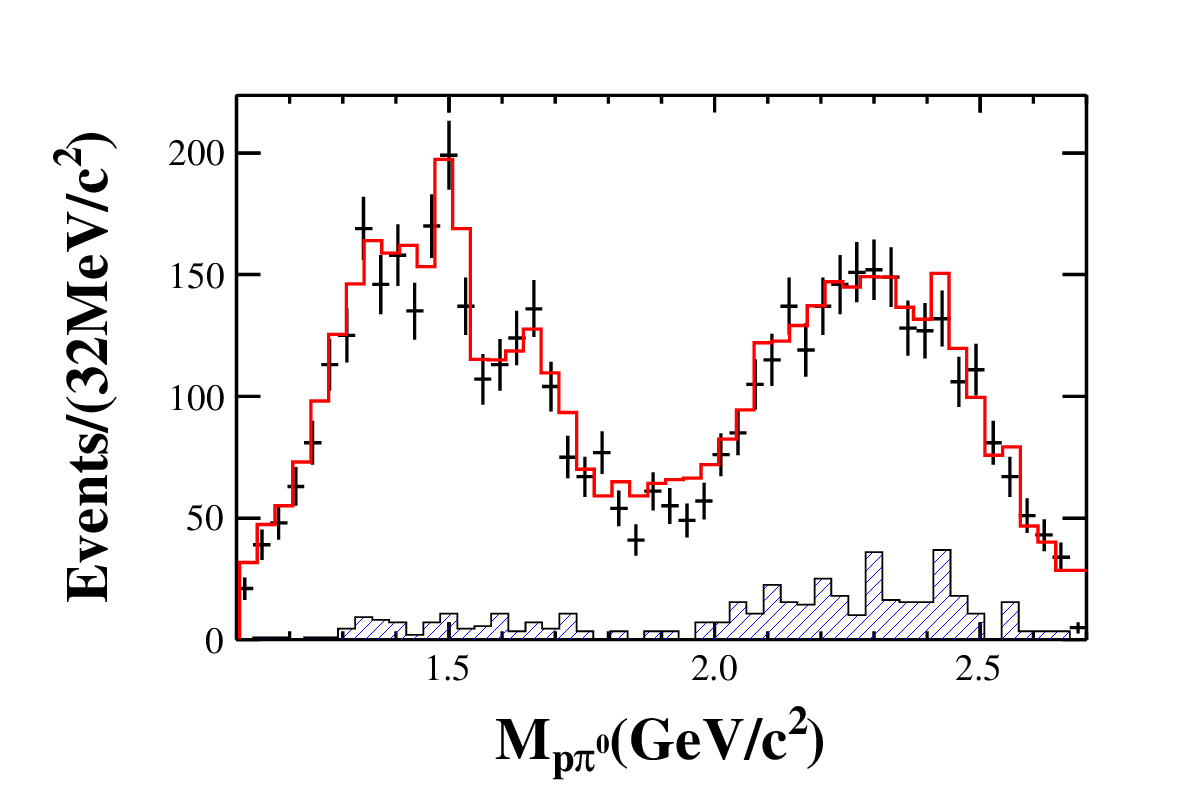,width=4.6cm, angle=0}
              \put(-27,67){(c)}
   \psfig{file=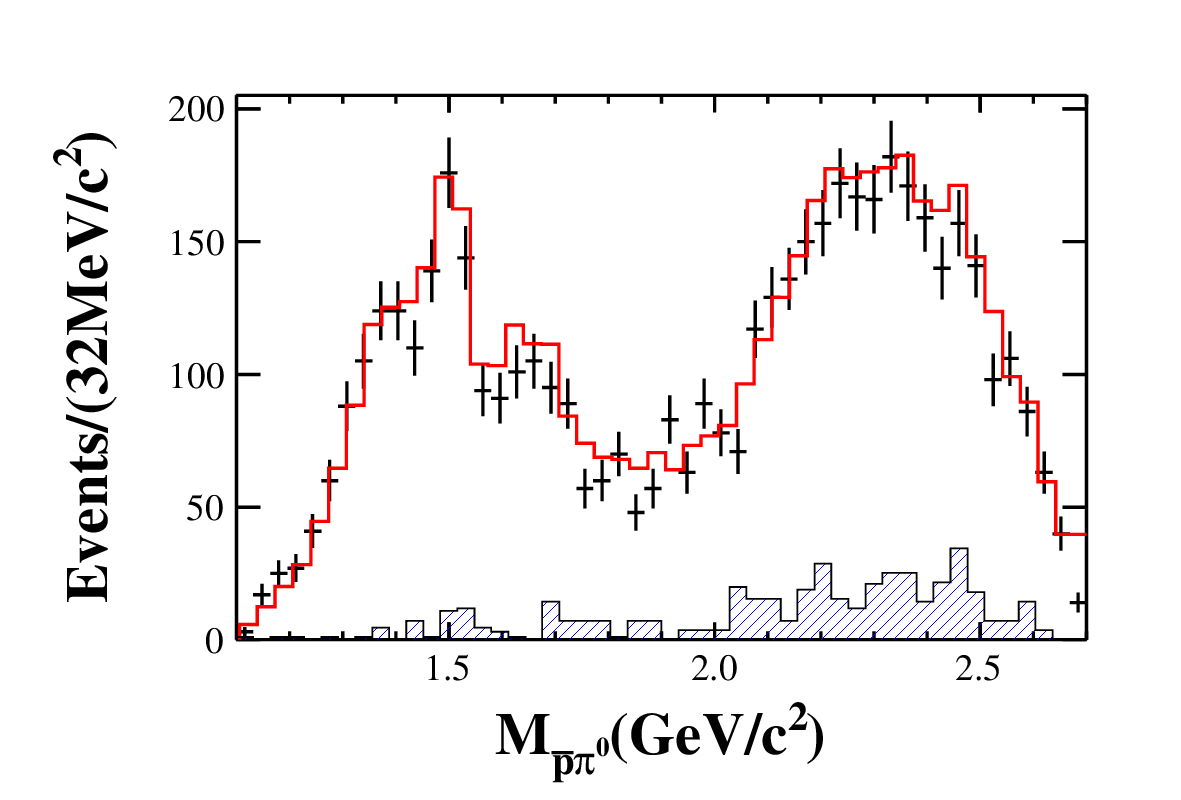,width=4.6cm, angle=0}
              \put(-27,67){(d)}}
            \caption{\label{fig:mppi0_signal} (a) Dalitz plot of
              $\psi(3686) \rightarrow p\bar{p}\pi^0$, the invariant
              mass spectra of (b) $p\bar{p}$, (c) $p\pi^0$, and (d)
              $\bar{p}\pi^0$. The dashed lines in (b) show the cut at
              the $J/\psi$ mass region. The crosses represent the
              experimental data, and the shaded histograms show the
              background from continuum process and $\pi^0$
              sideband. The histograms in solid line show the sum of
              MC prediction and the background. }

\end{figure}

The events passing the above selection criteria are shown in plots
(a) and (b) of Fig.~\ref{fig:mppi0_signal}, displayed as the Dalitz
plot of $\psi(3686) \rightarrow p\bar{p}\pi^0$ and the invariant
mass of $p\bar{p}$. The $p\bar{p}$ mass spectrum shows a clear
$J/\psi$ signal. Due to the detector resolution, the observed width
of $J/\psi$ is far larger than its natural width. This width
difference causes a problem in the inclusion of $J/\psi$ in partial
wave analysis. Thus, a cut of $|M_{p\bar{p}}-M_{J/\psi}|>$ 40
$\mathrm{MeV}/c^2$ is applied to exclude events with $p\bar{p}$
arising from $J/\psi$ decay. 4988 events survive the event selection
criteria. The mass spectra of $p\pi^0$ and $\bar{p}\pi^0$ for the
surviving events are shown in Fig.~\ref{fig:mppi0_signal} (c) and
(d).

For this analysis, two background sources are studied. The first one
arises from $\psi(3686)$ decays and has been studied with two
methods. In the first method, a sample of $10^8$ Monte Carlo
(MC)-simulated $\psi(3686)$ events is used and forty events survive
the event selection, mainly due to misidentified or lost photons. In
the second method, the background contribution is estimated using
the $\pi^0$ sideband events, defined by 30 $\mathrm{MeV}/c^2
<|M_{\gamma\gamma}-135|<$ 45 $\mathrm{MeV}/c^2$. Only 26 events are
found in the sideband area. The other background source arises from
the continuum process $e^+e^-\rightarrow \gamma*\rightarrow
p\bar{p}\pi^0$. This has been studied using 42 $pb^{-1}$ of
continuum data at $\sqrt{s} = 3650$ MeV. After normalizing to the
integrated luminosity of $\psi(3686)$, 447 background events are
found. In Fig.~\ref{fig:mppi0_signal}, the shaded histograms show
the total background contributions from continuum process and
$\pi^0$ sideband, in which the continuum contribution accounts for
about 95\%.

In our present investigation, with larger statistics than at CLEOc,
more than one $N^*$ state below 1700 MeV/$c^2$ is seen in the
$p\pi^0$ and $\bar{p}\pi^0$ mass spectra, and the threshold
enhancement in the $p\bar{p}$ mass spectrum is also visible. To
better understand the components of this decay, a partial wave
analysis taking into account the possible interferences is pursued.

The decay of $\psi(3686) \rightarrow p\bar{p}\pi^0$ is thought to be
dominated by two-body decays involving $N^*$, $\bar{N}^*$ states
~\cite{soft_pion_Nstar}, which can be described by
$\psi(3686)\rightarrow p\bar{N}^*(\bar{p}N^*)$,
$N^*(\bar{N}^*)\rightarrow p\pi^0(\bar{p}\pi^0)$. In addition, a
process of the type $\psi(3686)\rightarrow R\pi^0$ is considered,
where R represents a hypothetical $p\bar{p}$ resonance. The data are
fitted applying an unbinned maximum likelihood fit. The amplitudes
($A_i$) for all possible partial waves are constructed using the
relativistic covariant tensor amplitude
formalism~\cite{Ablikim:2009iw,
  amplitude_theory_1, amplitude_theory_2}. With these amplitudes, the
total transition probability for each event is obtained from a linear
combination of these partial wave amplitudes as $\omega =
|\Sigma_ic_iA_i|^2$. Finally the likelihood function $\ln(L)$ is
constructed as $\sum_{i=1}^{n} \ln(
\frac{\omega(\xi_i)\epsilon(\xi_i)}{\int
  d\xi\omega(\xi_i)\epsilon(\xi_i)})$, where n is the total number of
events, $\xi$ is the four-momentum of $p$, $\bar{p}$ and $\pi^0$,
$\omega(\xi)$ the probability density for a single event to populate
the phase space at $\xi$, and $\epsilon(\xi)$ is the detection
efficiency to detect one event with $\xi$. The free parameters $c_i$
are determined by maximizing the likelihood function $\ln(L)$. For
each $N^*$ state, the amplitude is parameterized with a Breit-Wigner
function, in which the mass and width of the resonance are
variables, as described in ~\cite{Ablikim:2009iw}. The background
contributions from $\pi^0$ sideband and continuum processes are
removed by subtracting the log-likelihood ($\ln(L)$) values, as the
log-likelihood value of data is the sum of that of signal and
background events. Possible interference between continuum processes
and $\psi(3686)$ decays is not considered.

All $N^*$ resonances up to 2200 MeV/$c^2$ with spin up to 5/2,
listed in the summary tables of the Particle Data Book~\cite{PDG},
are considered in this analysis, such as the well-established
states, $N(1440)$ and $N(1520)$, and not-well-measured states,
$N(2090)$ and $N(2100)$. Phase space decay and two speculative $N^*$
resonances, $N(1885)$ and $N(2065)$ are also considered. According
to the framework of soft $\pi$ meson theory~\cite{soft_pion}, the
off-shell decay process is needed in this channel. Thus, $N(940)$
with a mass of 940 MeV/$c^2$ and zero width is included. The
$N(940)$ represents a virtual proton, which could emit a $\pi^0$.
The Feynman diagram of this process can be found in
~\cite{Ablikim:2009iw}. In total, nineteen intermediate resonances
are considered.

For $N^*$ resonances with spin larger than 5/2, such as the
$N(2190)$, $N(2220)$, $N(2250)$ and $N(2600)$ ~\cite{High_spin_1,
High_spin_2, High_spin_3}, orbital angular momenta L$\ge$2 are
required, and are not expected to contribute significantly in
charmonium decay due to the suppression by the centrifugal barrier.
The reason is two-fold. At first, the annihilation radius of
$c\bar{c}$ is very small, estimated to be in the order of 0.1 fm,
due to the large mass of charm quark. This is about one order of
magnitude smaller than the interaction radius of $\pi N$ scattering
which is about several fm. Secondly, the relative momentum of $N^*$
and $\bar{p}$ is small, especially for large mass $N^*$ resonance.
Given the small annihilation radius and the small relative momenta
of $N^*$ and $\bar{N}$, orbital angular momenta L$\ge$2 should be
suppressed. If otherwise high spin states do exist in this decay,
this should result in an inconsistency of data and fit, which is not
observed. Thus, with the sensitivity of the present experiment, we
consider it adequate to include only states with spin up to 5/2.

In our analysis, the first step is to select the significant
resonances among all these resonances. The significance of each
resonance is determined from the difference of the likelihood values
of fits with and without the given resonance, accounting for the
change of the number of parameters. Resonances with significance
greater than $5\sigma$ are taken as significant ones and include
$N(940)$ and seven $N^*$ resonances.  The remaining insignificant
resonances are removed and only considered when estimating the
systematic errors. The mass and width of $N^*$ states are varied,
and the values with the best fitting result are taken as the
optimized values. Table ~\ref{tab:mass_width_optimized} lists the
optimized values for the seven $N^*$ states. Here, the first errors
are statistical and the second ones are systematic. In this table,
the first five $N^*$ resonances are consistent with the values in
the Particle Data Book ~\cite{PDG}, while the last two states can
not be identified with $N(2100)$ or $N(2200)$. However, the
significance of these two states are 15$\sigma$ and 11.7 $\sigma$,
respectively. As a consequence, we label these two states as
$N(2300)$ and $N(2570)$, with $J^P$ assignment of $1/2^+$ and
$5/2^-$, respectively.

\begin{figure}[htbp]


   \centerline{
   \psfig{file=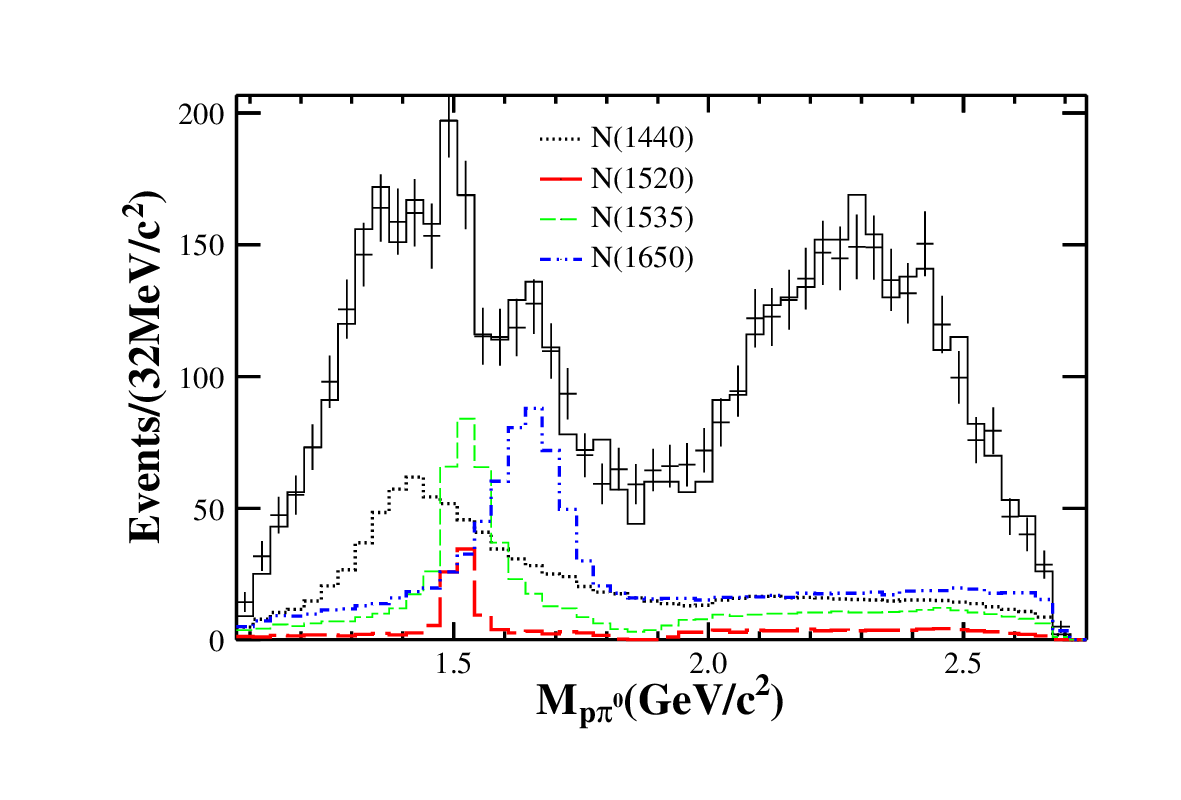,width=9cm, angle=0}
              \put(-40,145){(a)}}
   \centerline{
   \psfig{file=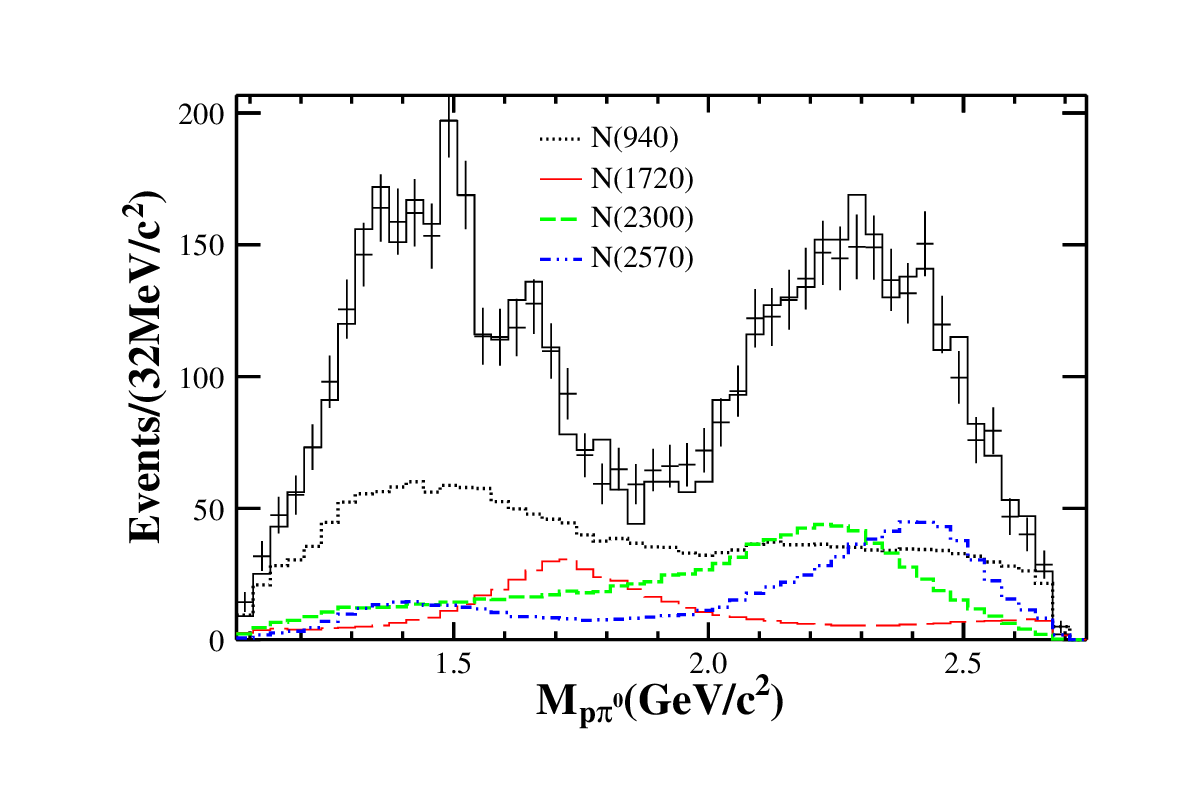,width=9cm, angle=0}
              \put(-40,145){(b)}}

   \caption{\label{fig:component} The contribution of each intermediate
resonance in the $p\pi^0$ mass spectra. The interferences with other
resonances are included.}

\end{figure}

Using these eight significant resonances, the fit result agrees well
with the data, as shown in Fig.~\ref{fig:mppi0_signal}. The $\chi^2$
over the number of degree of freedom is 1.12. The contribution of
each intermediate resonance including interference effects with
other resonances are extracted and shown in
Fig.~\ref{fig:component}. Plot (a) shows the contributions of
$N(1440)$, $N(1520)$, $N(1535)$ and $N(1650)$ in which we can see
clear peaks and also tails at the high mass region from the
interference effects. Plot (b) shows the contributions of $N(940)$,
$N(1720)$, $N(2300)$ and $N(2570)$. For $N(2300)$ and $N(2570)$,
their peak positions are below the Breit-Wigner mean values reported
in Table I because of the presence of interference contributions, as
well as phase space and centrifugal barrier factors.

\begin{table}[hbtp]
  \centering
  \footnotesize
  \vspace{-0.0cm} \caption{The optimized mass, width and significance
(Sig.) of the seven significant $N^*$ resonances. $\Delta S$
represents the change of the log likelihood value. $\Delta N_{dof}$
is the change of the number of free parameters in the fit. In the
second and third columns, the first error is statistical and the
second is systematic. The names of the last two resonances,
$N(2100)$ and $N(2200)$, have been changed to $N(2300)$ and
$N(2570)$ according to the optimized
masses.}\label{tab:mass_width_optimized}
\vspace{0.3cm}

\begin{ruledtabular}
\begin{tabular}{cccccc}

  Resonance & M(MeV/$c^2$) &  $\Gamma$(MeV/$c^2$) & $\Delta S$& $\Delta N_{dof}$ &  Sig.\\
\hline
$N(1440)$&$1390^{+11}_{-21}$$^{+21}_{-30}$&$340^{+46}_{-40}$$^{+70}_{-156}$
&72.5 & 4& $11.5\sigma$
\\ $N(1520)$& $1510^{+3}_{-7}$$^{+11}_{-9}$& $115^{+20}_{-15}$$^{+0}_{-40}$& 19.8 & 6 &$5.0\sigma$
\\ $N(1535)$& $1535^{+9}_{-8}$$^{+15}_{-22}$& $120^{+20}_{-20}$$^{+0}_{-42}$&49.4&4&$9.3\sigma$
\\ $N(1650)$& $1650^{+5}_{-5}$$^{+11}_{-30}$& $150^{+21}_{-22}$$^{+14}_{-50}$&82.1&4&$12.2\sigma$
\\ $N(1720)$& $1700^{+30}_{-28}$$^{+32}_{-35}$& $450^{+109}_{-94}$$^{+149}_{-44}$&55.6&6&$9.6\sigma$
\\ $\bf{}N(2300)$& $\bf{}2300^{+40}_{-30}$$\bf{}^{+109}_{-0}$&$\bf{}340^{+30}_{-30}$$\bf{}^{+110}_{-58}$&\bf{}120.7&\bf{}4&$\bf{}15.0\sigma$
\\ $\bf{}N(2570)$& $\bf{}2570^{+19}_{-10}$$\bf{}^{+34}_{-10}$ & $\bf{}250^{+14}_{-24}$$\bf{}^{+69}_{-21}$&\bf{}78.9&\bf{}6&$\bf{}11.7\sigma$
\end{tabular}
\end{ruledtabular}
\end{table}


Various checks have been performed to test the reliability of this
analysis. The first one is the spin parity check, in which the spin
parity of each state of the optimized solution is changed to other
possible values to test the other $J^P$ assignments. For $N(2300)$
and $N(2570)$, $1/2^+$ and $5/2^-$, respectively, are the best $J^P$
values. The significance becomes worse using other $J^P$
assignments. The second one is the Input-Output check. A MC sample
was generated with given components. After the fitting procedure
described above, the significant states and their properties (mass,
width, branching fraction, and the effect of interference terms) are
compared with the input values. The output values agree with the
input within $\pm 1\sigma$, corroborating that the analysis
procedure is reliable.

On the basis of the eight significant states, a scan for additional
resonances has been performed with different spin parity, mass and
width combinations. No extra resonance has been found to be
significant. For $N(1885)$, the obtained significance ranges from
1$\sigma$ to 1.2$\sigma$ depending on the mass and width. The
largest significance is obtained at a mass of 1930 MeV/$c^2$ and
width of 150 MeV/$c^2$. The significance for $N(2065)$ varies
between 3.2$\sigma$ and 4$\sigma$, where the maximum is obtained at
a mass of 2140 MeV/$c^2$ and width of 250 MeV/$c^2$. We consider
neither resonance as significant and do not claim any evidence.
Besides the known and speculative $N^*$ resonances, a $1^{--}$
$p\bar{p}$ resonance candidate described by the Breit-Wigner
function has been added, as suggested by the near-threshold
enhancement in the $p\bar{p}$ mass distribution. Varying the width
from 50 MeV/$c^2$ to 300 MeV/$c^2$ and mass from 1800 MeV/$c^2$ to
3000 MeV/$c^2$ with the step size of 10 MeV/$c^2$, the largest
significance obtained is 4$\sigma$ at a mass of 2000 MeV/$c^2$ and
width of 50 MeV/$c^2$, indicating that no $p\bar{p}$ resonance is
required to explain the threshold enhancement.

The branching fraction of $\psi(3686)\rightarrow p\bar{p}\pi^0$ is
determined as follows,

$$ B(\psi(3686)\rightarrow p\bar{p}\pi^0) = \frac{N-N_{bkg}}{\epsilon\times N_{\psi(3686)} \times B(\pi^0\rightarrow \gamma\gamma)}$$
$$ = (1.65\pm0.03\pm0.15)\times 10^{-4}$$

Here, $N$ represents the number of observed events, $N_{bkg}$ stands
for the number of estimated background events, and $\epsilon$ is the
efficiency derived from MC events generated according to the model
derived from the PWA analysis. This result is in agreement with the
value of $(1.33\pm0.17)\times 10^{-4}$ in the Particle Data Book
~\cite{PDG}. The products of the production and decay branching
fractions for each $N^*$ intermediate resonance are also determined,
as shown in Table ~\ref{tab:branching_fraction}. The sum of the
individual branching fractions is larger than the total due to
interference effects of the intermediate resonances.

\begin{table}[hbtp]
  \centering
  \footnotesize
  \vspace{-0.0cm}
\caption{Summary of measurements of the number of events, the MC
efficiency($\epsilon$), and the branching fraction (B.F.) of each
intermediate resonance and the whole channel. Here, for the number
of events and the branching fraction, the first error is statistical
and the second is systematic.}\label{tab:branching_fraction}

\vspace{0.3cm}
\begin{ruledtabular}
\begin{tabular}{cccc}
 Resonance &  N & $\epsilon(\%)$ & B.F.($\times10^{-5}$)\\
 \hline
$N(940)$  & $1870^{+90}_{-90}$$^{+487}_{-327} $& $27.5\pm0.4$ &
$6.42^{+0.20}_{-0.20}$$^{+1.78}_{-1.28}$
\\ $N(1440)$& $ 1060^{+90}_{-90}$$^{+459}_{-227}$ & $27.9\pm0.4$ &
$3.58^{+0.25}_{-0.25}$$^{+1.59}_{-0.84}$
\\ $N(1520)$& $ 190^{+14}_{-14}$$^{+64}_{-48}$  & $28.0\pm0.4$ &
$0.64^{+0.05}_{-0.05}$$^{+0.22}_{-0.17}$
\\ $N(1535)$& $ 673^{+45}_{-45}$$^{+263}_{-256}$  & $25.8\pm0.4$ &
$2.47^{+0.28}_{-0.28}$$^{+0.99}_{-0.97}$
\\ $N(1650)$& $ 1080^{+77}_{-77}$$^{+382}_{-467}$ & $27.2\pm0.4$ &
$3.76^{+0.28}_{-0.28}$$^{+1.37}_{-1.66}$
\\ $N(1720)$& $ 510^{+27}_{-27}$$^{+50}_{-197}$ & $26.9\pm0.4$ &
$1.79^{+0.10}_{-0.10}$$^{+0.24}_{-0.71}$
\\ $N(2300)$& $ 948^{+68}_{-68}$$^{+394}_{-213}$  & $34.2\pm0.4$ &
$2.62^{+0.28}_{-0.28}$$^{+1.12}_{-0.64}$
\\ $N(2570)$& $ 795^{+45}_{-45}$$^{+127}_{-83}$  & $35.3\pm0.4$ &
$2.13^{+0.08}_{-0.08}$$^{+0.40}_{-0.30}$
\\ Total& 4515$\pm$93 & $25.8\pm0.4$ & $16.5\pm0.3\pm1.5$

\end{tabular}
\end{ruledtabular}

\end{table}

The systematic uncertainty sources are divided into two categories.
The first includes the systematic errors from the number of
$\psi(3686)$ events (4\%), MDC tracking (4\% for two charged
tracks), particle identification (2\% for both proton and
anti-proton), photon detection efficiency (2\%), and kinematic fit
(7\%). These uncertainties are applicable to all branching fraction
measurements. The total systematic error from these common sources
is 9.4\%. The second source concerns the fitting procedure, which
includes the uncertainties from additional possible resonances, the
uncertainties using different Breit-Wigner parameterizations for
partial wave amplitude, the uncertainties from background
estimation, the uncertainties from the $J/\psi$ exclusion cut, as
well as the differences in the Input-Output check. These sources are
applied to the mass, width and branching fraction measurements of
intermediate states. The total systematic errors are the combination
of the errors from the common sources and the fitting procedure.


In summary, we studied the intermediate resonances, including their
masses, widths and spin parities, in the decay
$\psi(3686)\rightarrow p\bar{p}\pi^0$. Two new $N^*$ resonances are
observed, in addition to five well-known $N^*$ resonances. The
masses and widths as well as the spin parities of the two new $N^*$
states have been measured.  The branching fractions of
$\psi(3686)\rightarrow p\bar{p}\pi^0$ and the product branching
fractions through each intermediate $N^*$ state are measured. No
clear evidence for $N(1885)$ or $N(2065)$ has been found. The
hypothetical $p\bar{p}$ resonance has a significance of less than
4$\sigma$, indicating that the threshold enhancement most likely is
due to interference of $N^*$ intermediate resonances.




The BESIII collaboration thanks the staff of BEPCII and the
computing center for their hard efforts. This work is supported in
part by the Ministry of Science and Technology of China under
Contract No. 2009CB825200; National Natural Science Foundation of
China (NSFC) under Contracts Nos. 10625524, 10821063, 10825524,
10835001, 10935007, 11125525; Joint Funds of the National Natural
Science Foundation of China under Contracts Nos. 11079008, 11179007;
the Chinese Academy of Sciences (CAS) Large-Scale Scientific
Facility Program; CAS under Contracts Nos. KJCX2-YW-N29,
KJCX2-YW-N45; 100 Talents Program of CAS; Istituto Nazionale di
Fisica Nucleare, Italy; U. S. Department of Energy under Contracts
Nos. DE-FG02-04ER41291, DE-FG02-91ER40682, DE-FG02-94ER40823; U.S.
National Science Foundation; University of Groningen (RuG);
Helmholtzzentrum fuer Schwerionenforschung GmbH (GSI), Darmstadt;
WCU Program of National Research Foundation of Korea under Contract
No. R32-2008-000-10155-0.

\end{document}